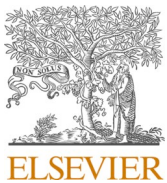
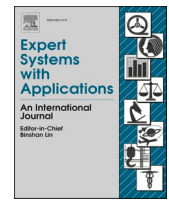

# Multi-stage transfer learning for lung segmentation using portable X-ray devices for patients with COVID-19

Plácido L. Vidal [*,1], Joaquim de Moura [2], Jorge Novo [3], Marcos Ortega [4]

*Centro de investigación CITIC, Universidade da Coruña, Campus de Elviña, s/n, 15071 A Coruña, Spain*
*Grupo VARPA, Instituto de Investigación Biomédica de A Coruña (INIBIC), Universidade da Coruña, Xubias de Arriba, 84, 15006 A Coruña, Spain*

ABSTRACT

One of the main challenges in times of sanitary emergency is to quickly develop computer aided diagnosis systems with a limited number of available samples due to the novelty, complexity of the case and the urgency of its implementation. This is the case during the current pandemic of COVID-19. This pathogen primarily infects the respiratory system of the afflicted, resulting in pneumonia and in a severe case of acute respiratory distress syndrome. This results in the formation of different pathological structures in the lungs that can be detected by the use of chest X-rays. Due to the overload of the health services, portable X-ray devices are recommended during the pandemic, preventing the spread of the disease. However, these devices entail different complications (such as capture quality) that, together with the subjectivity of the clinician, make the diagnostic process more difficult and suggest the necessity for computer-aided diagnosis methodologies despite the scarcity of samples available to do so. To solve this problem, we propose a methodology that allows to adapt the knowledge from a well-known domain with a high number of samples to a new domain with a significantly reduced number and greater complexity. We took advantage of a pre-trained segmentation model from brain magnetic resonance imaging of a unrelated pathology and performed two stages of knowledge transfer to obtain a robust system able to segment lung regions from portable X-ray devices despite the scarcity of samples and lesser quality. This way, our methodology obtained a satisfactory accuracy of $0.9761 \pm 0.0100$ for patients with COVID-19, $0.9801 \pm 0.0104$ for normal patients and $0.9769 \pm 0.0111$ for patients with pulmonary diseases with similar characteristics as COVID-19 (such as pneumonia) but not genuine COVID-19.

## 1. Introduction

The World Health Organization (WHO) declared a global health emergency on January 30th, 2020, due to the spread of SARS-CoV-2 and its disease COVID-19 beyond the People's Republic of China. Thus, the pandemic surpassed the million of deaths as well as tens of millions of people infected worldwide (Coronavirus Resource Center & Johns Hopkins, 2020).

For these reasons, at the dawn of the pandemic, proven computational methodologies of medical image analysis have been tested, as well as developing new ones with the aim of facilitating, accelerating and reducing the subjectivity factor of diagnostics at a critical moment for humanity (Shi et al., 2020; Shoeibi et al., 2020). Most of these methodologies are based on deep learning strategies, except for some particular proposals that use classic machine learning approaches (Hassanien, Mahdy, Ezzat, Elmousalami, & Ella, 2020) or others that actually use these techniques as support for deep learning methods (Mei et al., 2020; Sethy & Behera, 2020).

Regarding methodologies that aimed to help with the diagnostic of COVID-19 based on deep learning and convolutional neural networks (CNN), one of the first trends is to use these strategies to perform a medical screening. These methodologies return a label or severity of a COVID-19 candidate patient (Islam, Islam, & Asraf, 2020; Ozturk et al., 2020; de Moura, Novo, & Ortega, 2020; de Moura et al., 2020; Zhang et al., 2020).

Other trend with deep learning automatic approaches is to aid in the segmentation of the pulmonary region of interest. This region, as mentioned, is hard to correctly assess due to the difficulties of analyzing a radiography (Joarder & Crundwell, 2009) but critical, as the COVID-19 clinical picture mainly manifests its effects in the lung parenchyma (even after the patient has been discharged (Mo et al., 2020)). These works are usually integrated as input of other methodologies to improve their results by reducing the search space to only the region of interest or as a mean to greatly improve a posterior visualization of these results (Yan et al., 2020).

The third trend consists in, instead of trying to segment these lung regions, as they tend to be obfuscated by other tissues in the chest region, try to directly obtain the pathological structures of COVID-19 (Fan et al., 2020).

And, finally, works that try to palliate or complement their approaches by merging some (or all) of the mentioned trends into a single methodology (Alom, Rahman, Nasrin, Taha, & Asari, 2020; Chen, Yao, & Zhang, 2020).

Our work aims at following the second paradigm, extracting the lung regions, but specifically for images that are captured by portable X-ray devices. These devices present lower capture quality and, therefore, higher complexity. To the best of our knowledge, there are no other systems specially designed to work with chest radiographs obtained from these portable machines. This is specially relevant as these devices are recommended by the American College of Radiology (ACR) during emergency situations because they help to minimize the risk of cross-infection and allow for a comfortable and flexible imaging of the patients (American College of Radiology, 2020). In addition, these systems are ideal for emergency and saturation prevention of the healthcare services, as they do not require strict structuring of the established circuit and protocol (Jacobi, Chung, Bernheim, & Eber, 2020; Wong et al., 2020). A comparative summary of all the aforementioned proposals against ours can be seen in Table 1.

As an example, Fig. 1 shows three representative images from clinical practice with these portable devices for three different cases: patients with diagnosed COVID-19, patients with pathologies unrelated to COVID-19 but with similar impact in the lungs, and normal lungs. These images show how the images that are taken with these portable devices tend to blur the tissues of the lung region, as well as the pathological

**Table 1**
Summary of some representative works of the state of the art in comparison with our proposal. As shown, none of them work in lung chest segmentation of images from portable devices and are able to work with a significantly limited dataset.

| Author | Objective | Strategy | Image types |
| --- | --- | --- | --- |
| Hassanien et al. (2020) | COVID-19 detection | Thresholding Support Vector Machine | General purpose chest radiographs |
| Mei et al. (2020) | COVID-19 detection | Combination of two CNN and an SVM to generate a joint model | Computerized Tomography Images |
| Sethy and Behera (2020) | COVID-19 detection | Resnet50 to extract deep features and SVM for classification | General purpose chest radiographs |
| Apostolopoulos and Mpesiana (2020) | COVID-19 classification (Normal, COVID-19, Pneumonia) | Transfer learning from generic domains | General purpose chest radiographs |
| Islam et al. (2020) | COVID-19 classification (Normal, COVID-19, Pneumonia) | CNN + Long Short-Term Memory networks | General purpose chest radiographs |
| Rahimzadeh and Attar (2020) | COVID-19 classification (Normal, COVID-19, Pneumonia) | Concatenation of two CNNs | General purpose chest radiographs |
| Loey et al. (2020) | COVID-19 classification (COVID-19, normal, bacterial pneumonia, and viral pneumonia) with a limited dataset | Generative Adversarial Networks and Transfer Learning | General purpose chest radiographs |
| Ucar and Korkmaz (2020) | COVID-19 classification (Normal, COVID-19, Pneumonia) | Bayesian-optimized CNN | General purpose chest radiographs |
| Ozturk et al. (2020) | COVID-19 classification (No, COVID-19, Pneumonia) | You Only Look Once object detection with DarkNet CNN | General purpose chest radiographs |
| de Moura et al. (2020) | COVID-19 classification (Normal, COVID-19, Pneumonia) | U-Net CNN | General purpose chest radiographs |
| de Moura et al. (2020) | COVID-19 classification (Normal, COVID-19, Pneumonia) | U-Net CNN | Chest radiographs exclusively from portable devices |
| Yan et al. (2020) | Segmentation of lung and COVID-19 regions in CT images | U-Net CNN | Computerized Tomography Images |
| Fan et al. (2020) | COVID-19 lung infection segmentation | Inf-Net CNN | Computerized Tomography Images |
| Chen et al. (2020) | Lung and COVID-19 infected regions segmentation in CT | Residual Attention U-Net | Computerized Tomography Images |
| Zhang et al. (2020) | COVID-19 detection | Anomaly detection with a Confidence-aware anomaly detection CNN | General purpose chest radiographs |
| Alom et al. (2020) | COVID-19 detection, lung segmentation and infected region localization | Inception Recurrent Residual Neural Network for detection and NABLA-N network for segmentation | General purpose chest radiographs |
| Ours | Lung region segmentation (Normal, COVID-19, Pneumonia) with a limited dataset and poor image quality | Two-stage transfer learning from MRI glioma segmentation to general purpose chest radiographs to portable device chest radiographs with U-Net CNN | Chest radiographs exclusively from portable devices |

artifacts (specially in the images from afflicted lungs).

One of the first and most prominent symptoms of COVID-19 is the development of viral pneumonia, highlighting fever, cough, nasal congestion, fatigue, and other respiratory tract related affections (Velavan & Meyer, 2020). These symptoms manifest themselves in the lungs as ground glass abnormalities, patchy consolidations, alveolar exudates and interlobular involvement (Garg, Prabhakar, Gulati, Agarwal, & Dhooria, 2019; Brunese, Mercaldo, Reginelli, & Santone, 2020).

On the one hand, the ground glass abnormalities in chest radiographs are seen as diffuse translucent homogeneous brighter regions than the normal dark background of the lungs (albeit with a dim tone, nonetheless) usually caused by an inflammation of the tissues by the viral infection. On the other hand, the patchy consolidations are present as an irregular bright lattice pattern that could reach an homogeneous texture if the disease is quite advanced. These structures appear when regions of the lungs are filled with foreign fluids instead of normal air that alter the density. An example of these two cases can be seen in Fig. 2. In more serious cases, the patients may present acute respiratory distress syndrome or even systemic symptomatic manifestations (Gavriatopoulou et al., 2020; Lodigiani et al., 2020; Zaim, Chong, Sankaranarayanan, & Harky, 2020).

Performing a diagnostic with these portable devices is particularly challenging, as the generated images are of lesser quality due to the capture conditions, more difficult to inspect visually (as they usually only allow for a supine image instead of the usual multiple perspectives) and, due to the fact that they are obtained in emergencies, less available to researchers. For this reason, in this work we designed a segmentation methodology especially for images of low quality from portable devices and that is able to work with a limited number of samples. To the best of our knowledge, there is no other methodology specifically designed to analyze a set of images including COVID-19, also being taken in these particular challenging capture conditions and scarcity of samples.

To solve this issue, we developed a training methodology based on two stages of transfer learning between designed subsequent domains. Firstly, we took advantage of the knowledge learnt by a segmentation network from another medical imaging domain trained with a larger number of images and adapted it to be able to segment general lung chest images of high quality, including COVID-19 patients. Then, using a limited dataset composed by images from portable devices, we adapted the trained model from general lung chest X-ray segmentations to work specifically with images from these portable devices.

The proposal would allow to delimit the pulmonary region of interest, critical for the location of the pathological structures caused by COVID-19, independently from the subjectivity of the clinician (a subject particularly sensitive in situations of high stress and psychological wear) and under adverse capture conditions. Moreover, this system can be used as input to another methodology to reduce the search space to the lung region of interest or facilitate the subsequent visualization of the results.

In summary, the main contributions of this article are:

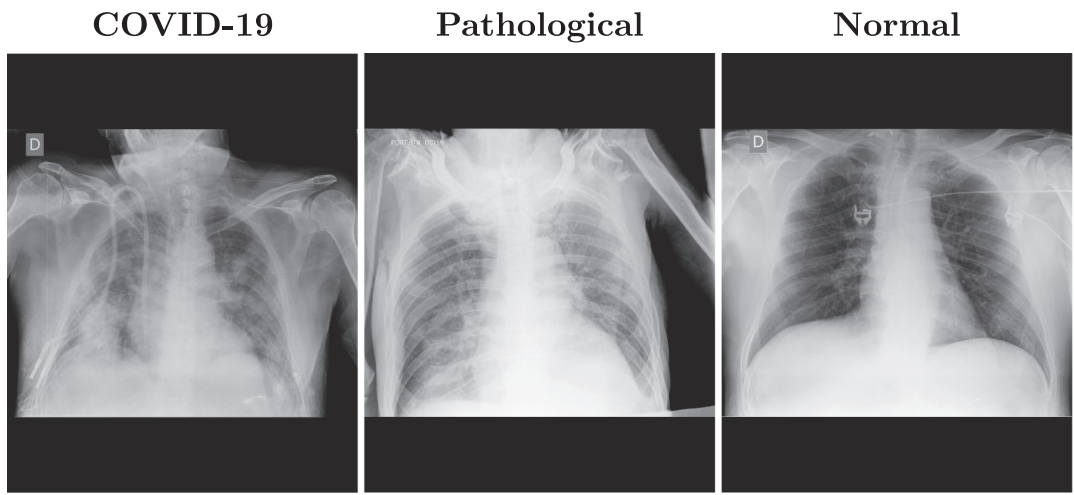

**Fig. 1.** Examples of images from portable devices for patients diagnosed with COVID-19, non-COVID-19 pathological lungs with similar characteristics and with normal lungs.

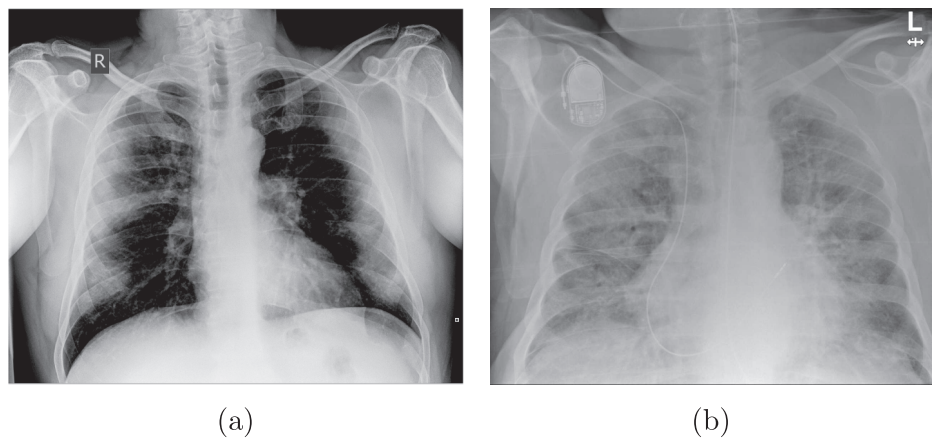

**Fig. 2.** Example of the main features present in COVID-19 lung radiographs. (a) Image with ground glass abnormalities. (b) Image with dense/lattice consolidations.

- Fully automatic proposal to segment the pulmonary region in low quality chest radiographs.
- Multiple stages of transfer learning between designed subsequent image domains to work with a limited number of portable X-ray samples.
- Datasets obtained from real clinical practice with portable devices (recommended when risk o cross-contamination and crowded hospital wings).
- To the best of our knowledge, our proposal represents the only fully automatic study specifically designed to work with portable capture devices.
- Robust and accurate even with poor quality images from these portable devices.
- Tested with images from COVID-19, pulmonary pathologies with similar characteristics to COVID-19 and normal lungs.

The present document is divided into six main sections. Section 2: "Materials", presents all the resources needed to fully reproduce our work in detail. Section 3: "Methodology" includes a complete explanation of the algorithm and strategy followed in this work and the particular parameters for each experiment. Section 4: "Results" presents the outcomes of the experiments described in Section 3 employing different metrics to evaluate different and complementary points of view. All these results are analyzed in Section 5: "Discussion", where we comment on different strengths, weaknesses and highlights of the methodology. Finally, Section 6: "Conclusions", which includes a final notes drawn for this research and a commentary on future lines of work.

## 2. Materials

Below, we will proceed to describe in detail the required materials and resources for the implementation and full reproduction of our work. In this section, the reader can see information and references of the datasets (Subsection 2.1), different software resources and their precise versions (Subsection 2.2) and hardware information as well as particular configuration of the equipment where the present project was conducted (Subsection 2.3).

### 2.1. Datasets

In this work, as we perform a double knowledge transfer, we need two different chest radiography datasets: the first one illustrating the general image domain and from which a larger number of samples are available (which we will call "General COVID lung dataset") and another dataset containing explicit samples from the target domain. This second dataset will contain images obtained in live clinical practice from a local hospital during the COVID-19 pandemic. Specifically, from the Universitary Hospital Complex of A Coruña (CHUAC, by its acronym in Spanish). For this reason, we will address this second dataset as the "CHUAC dataset". We will now proceed to explain in more detail the specifications and construction of each of the two datasets mentioned above.

#### 2.1.1. General COVID lung dataset

This first dataset was formed from public available datasets (Cohen et al., 2020; Kermany, 2018). The dataset contains images with varying resolutions, ranging from $5600 \times 4700$ pixels to $156 \times 156$ pixels including chest, lateral X-rays and CT images. For our purpose we discarded the latter two types. This was done because the portable devices of the consulted healthcare services were used only for chest X-rays. The dataset was labeled online in collaboration with different experts through the Darwin platform (V7 Labs, 2020) and is composed of 6,302 chest radiographs, from which 438 correspond to patients diagnosed with COVID-19, 4,262 with lung pathologies similar to COVID-19 and 1,602 belonging to patients who (in principle) do not suffer from any of the previously mentioned conditions (albeit they can be affected by other pathologies).

#### 2.1.2. CHUAC dataset (portable devices)

The second dataset was provided by the radiology service of the CHUAC from A Coruña, Galicia (Spain) obtained from two portable X-ray devices: an Agfa dr100E GE, and an Optima Rx200. For the acquisition procedure, the patient lies in a supine position and a single anterior-posterior projection is recorded. For this purpose, the X-ray tube is connected to a flexible arm that is extended over the patient to be exposed to a small dose of ionizing radiation, while an X-ray film holder or an image recording plate is placed under the patient to capture images of the interior of the chest. All the images were obtained after triage in live medical wings specially dedicated for the treatment and monitoring of patients suspected of being afflicted by the COVID-19. These images were captured during clinical practice and emergency healthcare services in the peak of the pandemic of 2020. This dataset contains 200 images of patients diagnosed with COVID-19, 200 images of patients with lung affections similar to (but not caused by) COVID-19 and 200 patients with, in principle, no pulmonary afflictions but that may be affected by other diseases, for a total of 600 images. The dataset contains images with varying resolutions, ranging from $1526 \times 1910$ pixels to $1523 \times 1904$ pixels. Due to the inherent limitations of portable capture devices, all images belong to patients in the supine position and an anterior-posterior projection is recorded.

All the data, before being received by anyone outside of the CHUAC radiology service staff, passed through a process of anonymization to protect the privacy of the individuals. Additionally, all the images were stored in private servers and security protocols in place, with restricted access only to personnel involved in this project. The protocols for this study have been reviewed by the hospital board and are contained in an agreement with the hospital management.

### 2.2. Software resources

Regarding the software resources, we have used Python 3.7.9 with Pytorch 1.6.0 (Paszke et al., 2019) and Scikit-Learn 0.23.2 (Pedregosa et al., 2011). Additionally, we used a pre-trained model from the work of Buda, Saha, and Mazurowski (2019) and Buda (2020b), trained with images from 110 patients for a total of 7858 images (Buda, 2020a). This network is an implementation of an U-Net (Ronneberger, Fischer, & Brox, 2015) dedicated to the identification of brain structures in magnetic resonance imaging or MRI (Buda et al., 2019). Specifically, the original network has been trained to detect gliomas, a type of brain tumor diagnosed mainly by this imaging modality (Forst, Nahed, Loeffler, & Batchelor, 2014; Buda et al., 2019), problematic that share similar characteristics to our case, which is herein exploited. The precise architecture of this network is presented in Fig. 3. As can be seen in the figure, the network used is based on an encoder-decoder architecture. While the encoder learns the relevant filters to abstract the important information and process it in the bottleneck, the decoder will gradually generate the target segmentation. This network is characterized by having the encoder and decoder connected in what are know as "skip-connections". These skip-connections allow to reuse information from the input and the encoder that would have been filtered in the bottleneck in the process of reconstruction/generation of the decoder, getting to produce more accurate representations. For this reason, this architecture is widely used in image analysis methodologies, especially in the field of medical imaging.

### 2.3. Hardware resources

Since tests were conducted to evaluate the performance of the methodology as well as its use of resources (and to allow full reproducibility of results), we include in Table 2, the full disclosure of the components, drivers and software that have been used throughout the realization of this work and may have influenced its performance.

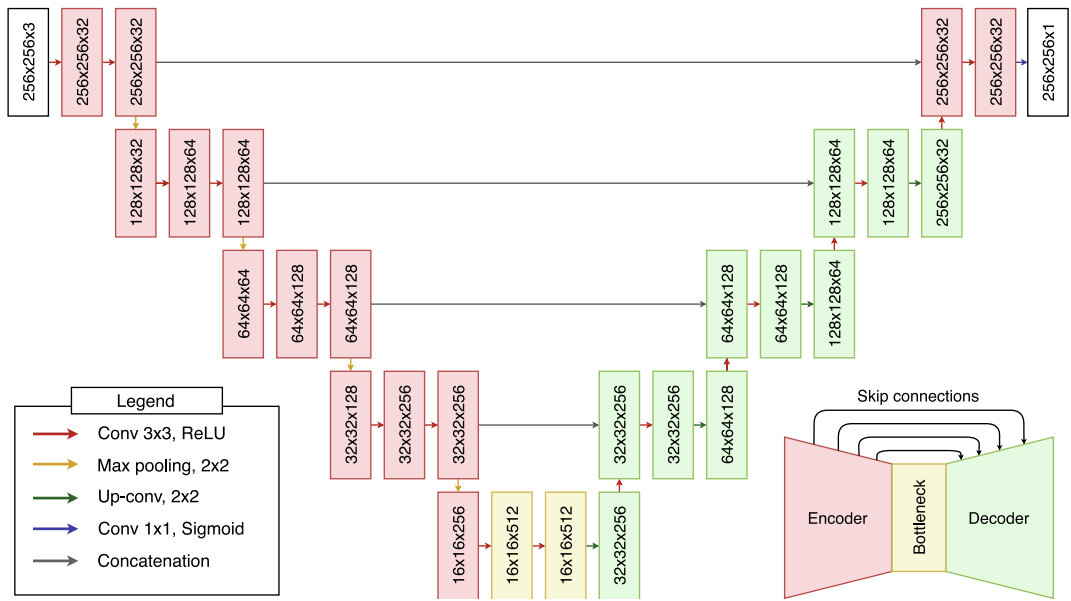

**Fig. 3.** Architecture of the pretrained convolutional neural network. Notice the encoder-decoder strategy with skip-connections, ideal for medical imaging segmentaton.

**Table 2**
Specifications of the equipment used throughout the project to carry out the experiments.

| Name | Description |
| --- | --- |
| OS | Ubuntu 18.04.5 LTS (Bionic Beaver) |
| Kernel | Linux 4.15.0–128-generic |
| Architecture | x86-64 |
| CPU | Intel(R) Core(TM) i9-9900 K CPU @ 3.60 GHz |
| Motherboard | ASUS PRIME Z390-A |
| RAM | 2 x 16GiB DIMM DDR4 Synchronous 2666 MHz CRUCIAL BLT16G4D26BFT4.C16FD |
| HDD | ATA Disk TOSHIBA DT01ACA2 1863GiB (2 TB) |
| GPU | NVIDIA Corporation GeForce RTX 2080 Ti |
| Driver Version | 450.51.06 |
| CUDA Version | 11.0 |

## 3. Methodology

To successfully develop a system able to work with radiographs from portable devices with a limited amount available from the saturated health services, we followed a workflow that allowed us to progressively adapt information from a different medical imaging domain and pathology to ours.

The main workflow followed in this work is detailed in Fig. 4, where we can see that each of the training stages performed in our project is repeated 25 times to evaluate the performance of the methodology. For each repetition, the dataset was be completely randomized and the variability of the test results analyzed to evaluate if more complex analytical and statistically robust strategies (such as cross-validation) were needed. In addition, the proposed fully automatic methodology was divided into two main stages of transfer learning. A first transfer learning stage to adapt the filters developed in the network for the MRI domain to chest radiography and a second one to further refine these weights specifically into the sensibly harder radiographs from portable devices. In the following two subsections, each stage will be explained in more detail. As both transfer learning stages share the same training algorithm, we will explain them together in Subsection 3.3: "Training details".

### 3.1. Inter domain knowledge transfer stage: MRI to common chest X-ray

For this first step, we started from a model previously trained in a medical imaging domain with a large and heterogeneous set of images that presents similar characteristics to those we would find in the target domain (from which we have available a limited number of samples). In our case, we used the U-Net trained with MRI images for glioma segmentation as shown in Section 2.2. As can be seen in Fig. 5, both image modalities present bright-to-dark diffuse gradients, dim lattice structures and sharp, steep formations with dark background (among others). Thus, while both pathologies are different in both origin and afflicted regions, a machine learning algorithm trying to analyze these image modalities must learn similar filters related to gradients, intensity and texture.

For this reason, the knowledge transfer between the two domains was direct. This was not only because of the similarity of characteristics of both image domains, but also because of the similar complications present in both image domains and pathologies. These factors made it an ideal candidate network to be the "knowledge donor" for our purpose.

This way, initially, we carried out a knowledge transfer stage by continuing the training of the network trained with a complete dataset of MRI images with general images of the domain to which we want to direct our work. Specifically, in this case, we have opted for the aforementioned public dataset. This dataset contains numerous radiographs obtained from different hospital and medical centers around the world (and, therefore, from a wide range of X-ray devices).

### 3.2. Inter device type knowledge transfer stage: common to X-ray images from portable devices

Once we had successfully trained a model to identify and segment lung regions in general chest X-ray images from patients with COVID-19, lung afflictions with similar characteristics or normal patients; we carried out the second stage of transfer learning. That is, we took advantage of the general patterns that the system has learned when looking for the pulmonary regions and we challenged them with images taken in adverse conditions to further refine the segmentations generated by the network. Consequently, when this second transfer learning stage was completed, we obtained a model specifically trained to search for pulmonary regions in the adverse conditions defined by the general dataset and our unfavorable dataset composed by radiographs taken with portable devices in adverse conditions.

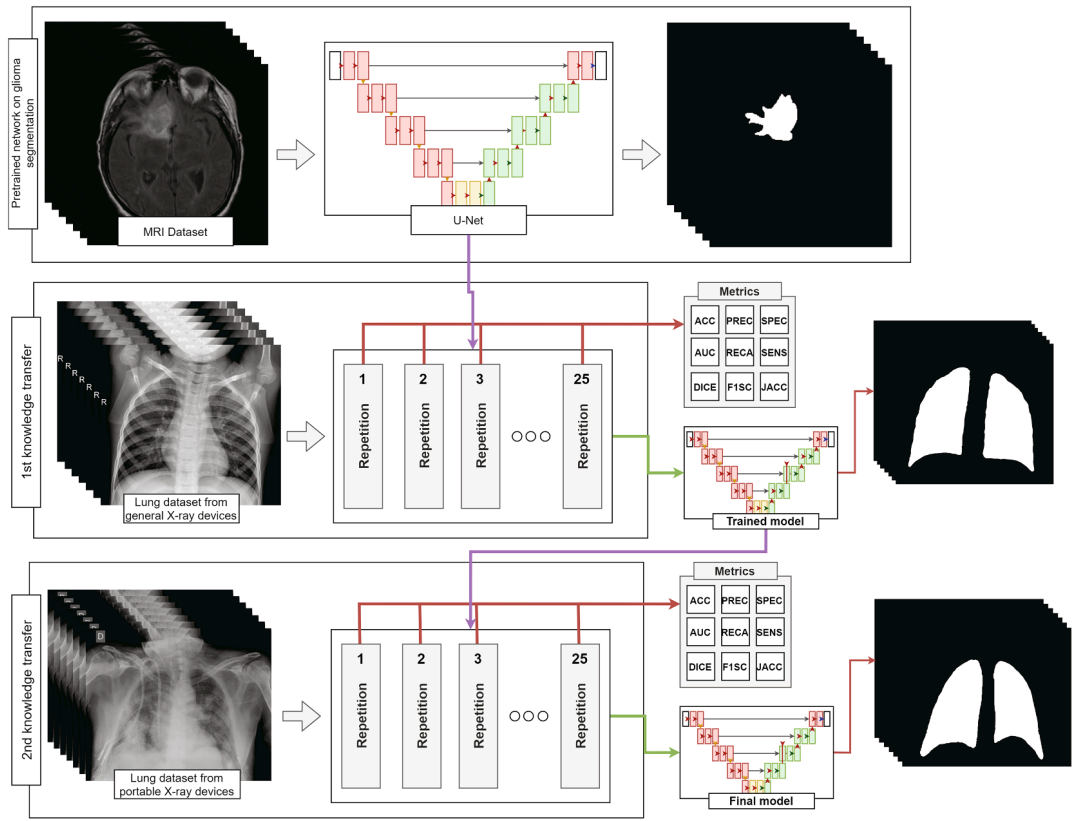

**Fig. 4.** Diagram of the fully automatic methodology to obtain a model able to segment lung regions in radiographs from portable devices.

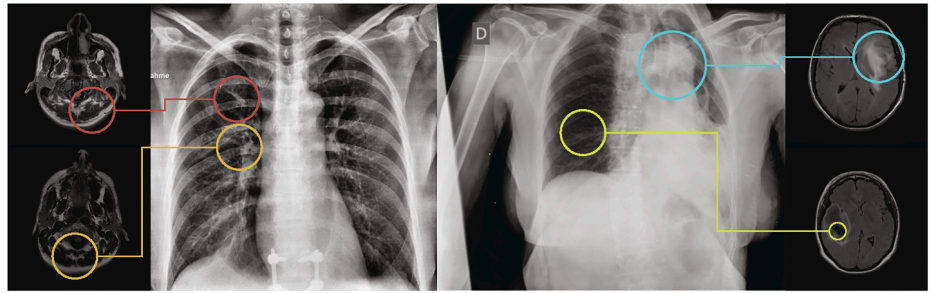

**Fig. 5.** Example MRI images and chest radiographs with gradient, texture and intensity patterns similar between domains.

In this stage, we also divided the dataset of 600 chest radiographs from portable X-ray devices obtained during clinical practice in the CHUAC into two datasets of 300 samples. This was done to use only one of the dataset halves to perform the knowledge transfer, and the other to evaluate the performance and improvement of the system before and after this stage.

### 3.3. Training details

In order to maintain consistency and allow for proper transfer learning, we have employed the same loss function used in the model trained with the brain MRI images for the subsequent transfer learning stages. Therefore, both models have been trained using the Smooth Dice Loss (Eq. (1)).

$$SmoothDiceLoss = 1 - 2\frac{|Op \cap Ot| + \lambda}{|Op| + |Ot| + \lambda} \quad (1)$$

where $Op$ represents the predicted system output and $Ot$ the expected output (target). $\lambda$ is the smoothing factor, which has been defined as 1 in this work. As optimizer, we have used adaptive moment estimation (ADAM) (Kingma & Ba, 2014), with a learning rate of 0.001 that is adjusted dynamically according to the necessities and training progression of the model. Finally, for the calculation of the number of training epochs we have used an early stopping strategy. That is, the algorithm will automatically stop when the model is not able to improve its performance. Specifically, the system evaluated the validation loss and had a patience of 20 epochs without obtaining any improvement. As for the distribution of the dataset, 60% of the samples have been used for the training of the model, 20% for the validation, and the remaining 20% for the unbiased testing of the model. Finally, as result of the training, the weights of the model of the network that obtained the best result in validation were recovered. This training process was repeated 25 times for a better evaluation of the training stages. Additionally, to increase the effective amount of available images in the dataset and to further improve the training, data augmentation techniques have been implemented. Specifically, the images were randomly rotated random degrees between $-10°$ and $+10°$ with a probability of 75% to simulate feasible postural variations of the patient. The detailed strategy followed for each training cycle is depicted in Fig. 6.

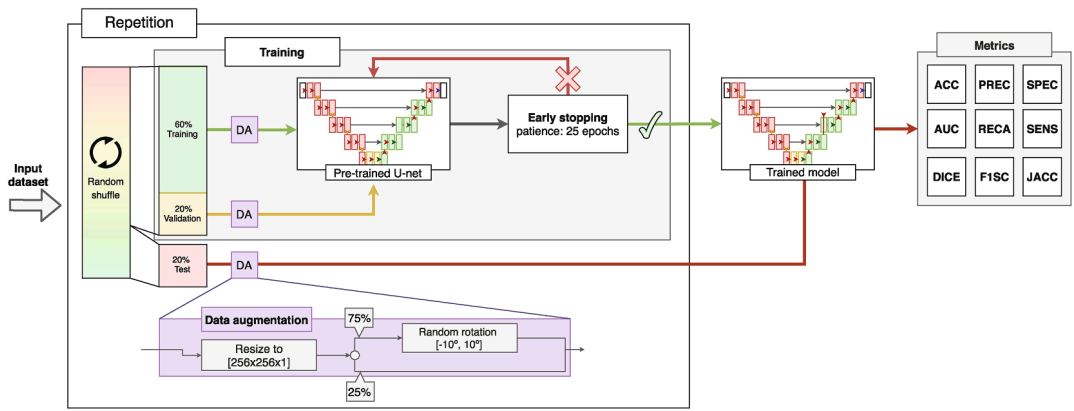

**Fig. 6.** Algorithm followed during each repetition, including the data augmentation process and early stopping configuration (the latter based on the validation error).

### 3.4. Evaluation

To evaluate the performance of our proposal in each stage, we analyzed the results in a wide range of metrics that allowed us to study the performance of each of the trained models from different points of view. To do so, we evaluated its accuracy (ACC), area under the ROC curve (AUC), Dice coefficient (DICE), Jaccard index (JACC), precision (PREC)[5], recall (RECA)[5], F1-Score (F1-SC)[5], sensitivity (SENS) and specificity (SPEC). In our case, and using as reference the True Positives (TP), True Negatives (TN), False Positives (FP). False Negatives (FN), $Ot$ as the target pixel values and $Op$ as the values predicted by the system for a given image, these metrics are defined as follows:[5]

$$ACC = \frac{TP + TN}{TP + TN + FP + FN} \quad (2)$$

$$DICE = 2 \times \frac{\sum(Ot \times Op)}{\sum Ot + \sum Op} \quad (3)$$

$$JACC = \frac{\sum(Ot \times Op)}{(\sum Ot + \sum Op) - \sum(Ot \times Op)} \quad (4)$$

$$PREC^5 = \frac{TP}{TP + FP} \quad (5)$$

$$RECA^5 = \frac{TP}{TP + FN} \quad (6)$$

$$F1 - SC^5 = 2 * \frac{(PREC \times RECA)}{(PREC + RECA)} \quad (7)$$

$$SENS = \frac{TP}{TP + FN} \quad (8)$$

$$SPEC = \frac{TN}{TN + FP} \quad (9)$$

Finally, AUC returns the probability that the analyzed model has of assigning a higher value to a positive sample over a negative sample (Bradley, 1997).

---

[5] PREC, RECA and F1-SC are a macro average: they are calculated for both the positive and negative classes and then averaged (thus, RECA and SENS display different values).

## 4. Results

In this section we will proceed to present the results that were obtained during the development of this work, product of the previously presented methodology.

### 4.1. Evaluation of the inter domain knowledge transfer stage: MRI to common chest X-ray

Now, we will proceed to present the results for the inter domain learning stage, where we took advantage of a model that was trained with a large number of images from a similar image domain (that allowed us to generate a robust methodology despite the scarcity of images available from portable devices). On this first stage, we adapted from this domain to common lung radiographs. The average of all the repetitions for this training process can be seen in Fig. 7, and the mean test results of each of the chosen models in Table 3. In this Fig. 7, we see that (on average) it does not need too many cycles to learn the patterns of the new domain, thanks to the already mentioned transfer of knowledge from similar modalities instead of starting the training from scratch.

As can be seen, thanks to the knowledge transfer stage we obtain a system capable of successfully segmenting the pulmonary region of interest. The only weak measurement is the one referring to the sensitivity of the model, with a considerably high standard deviation as well. However, the specificity obtained is considerably high, and with a very low deviation (which indicates consistency throughout the repetitions). These two factors indicate that the model is over-adjusting some of the detections. This is possibly due to the images showing foreign bodies such as pacemakers or other such objects, as the dataset masks (targets) have been corrected to try to estimate the complete lung surface even if it is obscured by these objects.

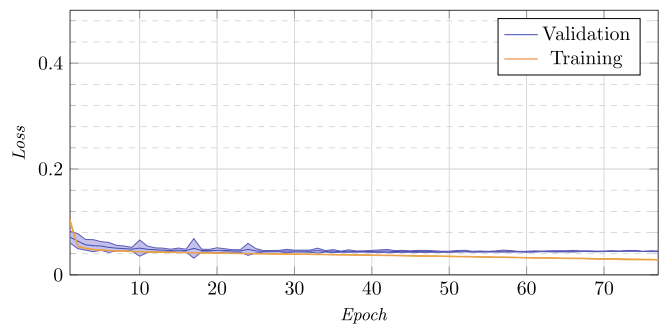

**Fig. 7.** Training and validation loss for the 25 repetitions for the inter domain knowledge transfer stage.

**Table 3**
Mean and standard deviation of test results for the 25 repetitions of the inter domain knowledge transfer stage.

|  | ACC | AUC | DICE | JACC | PREC | RECA | F1-SC | SENS | SPEC |
|---|---|---|---|---|---|---|---|---|---|
| Mean | 0.8813 | 0.9702 | 0.9554 | 0.9156 | 0.9064 | 0.8267 | 0.8366 | 0.6703 | 0.9832 |
| St. dev. | 0.0855 | 0.0244 | 0.0250 | 0.0432 | 0.0573 | 0.1247 | 0.1315 | 0.2528 | 0.0132 |

### 4.2. Evaluation of the inter device type knowledge transfer stage: Common to portable devices

After this first inter domain transfer learning, we now present the results of the inter device type transfer learning step. In this step, we used the model adapted for a general chest X-ray and continued the training to adapt this model to the final objective of this work: obtaining a robust system able to successfully segment lung regions in images taken in adverse conditions with portable devices.

In Fig. 8 we can see that, as in the inter domain learning stage, thanks to the use of image domains with similar properties, in just a few cycles we obtained the desired result. The graph can give the appearance of an slight over-training tendency, but we have to take into account two things: the first, that what is shown is the average of each epoch for 25 trainings, so the result shown is not really the training of a single model that shows a given behavior but multiple different behaviors averaged; the other is that we are dealing with a training that employs early stopping with 20 epochs of patience. The latter indicates that every model, in the same moment that they began to overtrain, automatically stopped the training and we were left with the best previous model. Despite the graph reaching more than 50 epochs, not all the models reached that many steps (and the further we go, the less models are affecting said mean, reflected in the standard deviation). Although the training decreases significantly compared to the validation error, the chosen model will not really present this pattern of training. Rather, what it indicates is that in early stages all models converge because they are based on an already-adapted model to the domain.

Finally, as can be seen in the test results of the chosen model in Table 4, the system appears to return more balanced results across all the metrics. We can see how the sensitivity of the system has sensibly improved and the system is now more balanced. Now, we will proceed to evaluate both systems under an unbiased dataset to better assess their differences and improvements.

#### 4.2.1. Evaluation of improvement between both knowledge transfer stages

For this final test we used the 300 independent images from the CHUAC dataset that we separated for further analysis. The results of these tests can be seen detailed in Tables 5 and 6; where we present the results for the test of the model before and after the second stage of transfer learning (where the model is adapted to portable X-ray devices), respectively.

Complementarily, this improvement is better observed in the comparison plots of Figs. 9–11. These graphs show that where the most noticeable change has been in images that have some kind of pathology with effects similar to COVID-19, improving by almost 0.02 points in Jaccard and DICE coefficients. On the other hand, we also noticed a remarkable increase in the sensitivity of the models, being this measurement critical in systems oriented to the medical sciences and clinical practice and also highly increased after the inter device type transfer learning stage into the portable X-ray image domain.

### 4.3. Evaluation of computational efficiency

Next, we present the results of an independent performance test with the same configuration used during training. These tests measure the time spent in milliseconds on the benchmark machine (Section 2). On average, the total time for a repetition (including memory operations and data conversions) consumed 2,332,775.82 ms on average with a standard deviation of 415,244.04 in the first knowledge transfer stage and 68,106.06 ms on average with a standard deviation of 13,286.57 in the second knowledge transfer stage. In the Table 7, you can see the time consumed in milliseconds during each epoch on average, both for the training and validation stages. All the epochs of the 25 repetitions were evaluated as one joined pool since each repetition takes an undetermined and variable number of epochs when using an early stopping strategy.

Finally, and exclusively measuring the network processing time of an image without taking into account any previous preprocessing and data transactions, the time required by the network to process each of the 300 images of the test set takes an average of 3.831 ms with a standard deviation of 0.286.

## 5. Discussion

In Fig. 12 we can see more closely examples of outputs generated by the proposed method. There is a clear distinction between the behavior of the two models. As we saw in the statistics, the model that was trained with a large number of MRI images and then its knowledge exploited to improve the training with the common lung radiographs tends to generate more adjusted and limited segmentations. This is particularly noticeable in those images that present less definition of the lung region, where the model would have to try to reconstruct the shape of the lung region on its own based on its inherited knowledge of the domain. On the other hand, the network that has been trained with the second stage of transfer learning presents more defined detections and better coverage of the lung regions.

However, the model resulting from the inter domain transfer learning stage also presents some explicit unwanted artifacts: it creates bridges between the lung regions and connected components isolated from the two lungs as can be seen in Fig. 13. In the same way, we see that in the final model from the inter device type transfer learning step all these artifacts have completely disappeared. Thus, we can see the reason behind the three phenomena observed in the comparison of the experiments: the overall improvement of the segmentations, the increase of the sensitivity and at the same time the reduction of the standard deviation of the metrics (as their stability is significantly improved with the disappearance of these unwanted artifacts from the inter domain transfer learning stage model).

Thanks to these comparisons we can see the advantage of applying our methodology based on two stages of transfer learning. In the images that only rely on the first stage we are simply seeing the performance of a model adapted to the general lung domain. However, after the application of the second transfer of knowledge we can see the effective performance gain of our work against proposals that only are competing with the first stage (as the results attained in the first stage are on par with the state of the art and use general chest radiographs). In this way, although our proposal is the first to work exclusively with this type of image, we are able to approximate what would be a fair comparison with other similar approaches to the general purpose lung segmentation state of the art, obtaining satisfactory results even with a limited number of samples and with images of lower quality.

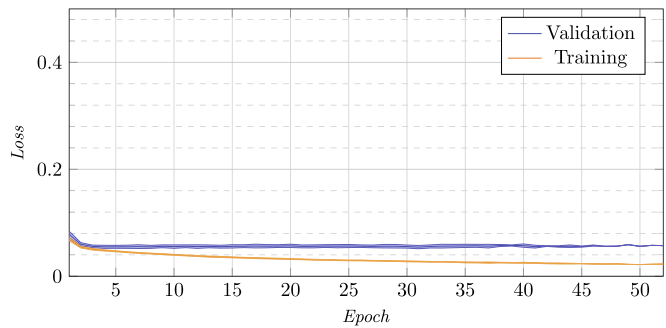

**Fig. 8.** Training and validation loss for the 25 repetitions for the inter device type knowledge transfer stage.

**Table 4**
Mean and standard deviation of test results for the 25 repetitions of the inter device type knowledge transfer stage.

|  | ACC | AUC | DICE | JACC | PREC | RECA | F1-SC | SENS | SPEC |
| --- | --- | --- | --- | --- | --- | --- | --- | --- | --- |
| Mean | 0.9773 | 0.9695 | 0.9436 | 0.8936 | 0.9656 | 0.9644 | 0.9641 | 0.9423 | 0.9864 |
| St. dev. | 0.0097 | 0.0243 | 0.0243 | 0.0419 | 0.0175 | 0.0228 | 0.0152 | 0.0500 | 0.0107 |

**Table 5**
Breakdown of the results from the model from the inter domain knowledge transfer stage tested with the CHUAC dataset and by pathology.

|  | ACC | AUC | DICE | JACC | PREC | RECA | F1-SC | SENS | SPEC |
| --- | --- | --- | --- | --- | --- | --- | --- | --- | --- |
|  |  |  |  |  | *COVID-19* |  |  |  |  |
| Mean | 0.9570 | 0.9377 | 0.8936 | 0.8142 | 0.9481 | 0.9286 | 0.9348 | 0.8729 | 0.9844 |
| St. dev. | 0.0293 | 0.0318 | 0.0698 | 0.1046 | 0.0471 | 0.0372 | 0.0418 | 0.0745 | 0.0230 |
|  |  |  |  |  | *Normal* |  |  |  |  |
| Mean | 0.9555 | 0.9326 | 0.8854 | 0.8014 | 0.9484 | 0.9220 | 0.9305 | 0.8576 | 0.9864 |
| St. Dev. | 0.0439 | 0.0288 | 0.0731 | 0.1073 | 0.0473 | 0.0493 | 0.0588 | 0.0973 | 0.0197 |
|  |  |  |  |  | *Pathological* |  |  |  |  |
| Mean | 0.9476 | 0.9228 | 0.8536 | 0.7551 | 0.9268 | 0.9145 | 0.9173 | 0.8536 | 0.9754 |
| St. dev. | 0.0294 | 0.0323 | 0.0928 | 0.1293 | 0.0584 | 0.0346 | 0.0468 | 0.0608 | 0.0286 |

**Table 6**
Breakdown of the results from the model from the inter device type transfer learning stage tested with the CHUAC dataset and by pathology.

|  | ACC | AUC | DICE | JACC | PREC | RECA | F1-SC | SENS | SPEC |
| --- | --- | --- | --- | --- | --- | --- | --- | --- | --- |
|  |  |  |  |  | *COVID-19* |  |  |  |  |
| Mean | 0.9761 | 0.9705 | 0.9447 | 0.8961 | 0.9653 | 0.9655 | 0.9644 | 0.9444 | 0.9867 |
| St. dev. | 0.0100 | 0.0204 | 0.0241 | 0.0411 | 0.0205 | 0.0193 | 0.0145 | 0.0443 | 0.0108 |
|  |  |  |  |  | *Normal* |  |  |  |  |
| Mean | 0.9801 | 0.9752 | 0.9528 | 0.9103 | 0.9724 | 0.9688 | 0.9701 | 0.9470 | 0.9906 |
| St. dev. | 0.0104 | 0.0158 | 0.0161 | 0.0288 | 0.0122 | 0.0171 | 0.0115 | 0.0373 | 0.0059 |
|  |  |  |  |  | *Pathological* |  |  |  |  |
| Mean | 0.9769 | 0.9649 | 0.9414 | 0.8910 | 0.9674 | 0.9616 | 0.9637 | 0.9340 | 0.9891 |
| St. dev. | 0.0111 | 0.0334 | 0.0322 | 0.0532 | 0.0184 | 0.0256 | 0.0189 | 0.0525 | 0.0077 |

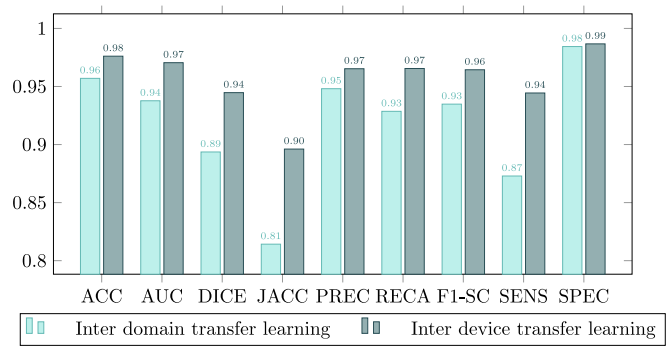

**Fig. 9.** Comparison between the model trained with common chest radiographs and the model adapted to portable devices for images from patients diagnosed with COVID-19.

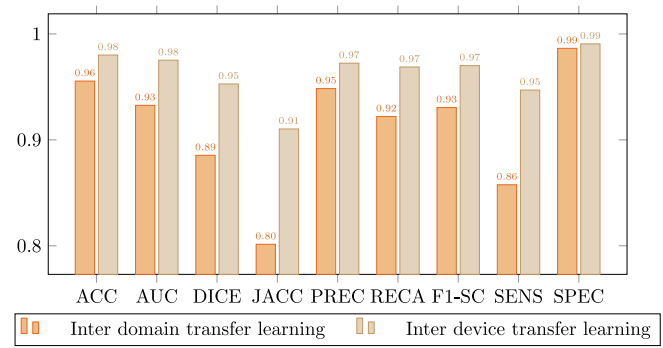

**Fig. 10.** Comparison between the model trained with common chest radiographs and the model adapted to portable devices for images from patients without lung afflictions.

In addition, by studying Figs. 7 and 8 and Tables 3 and 4 we can see that the randomized holdout division of the dataset repeated several times was enough to evaluate the performance of our methodology, since the statistical variability obtained in both during the training and testing of the model was insignificant and without any hint of imbalanced representation of the samples. Moreover, since we are testing with a higher number of repetitions compared to the usual 10 of the cross-validation, we are more than compensating any possible (and with negligible influence nonetheless on the results) bias that may have appeared in any iteration.

Another limitation of our methodology, manifested in the images as an slight loss in accuracy and smoothness in the borderline regions of the segmentations, relies in the rescaling of the images. All the input images in both phases of transfer learning are resized to the input size of the MRI-trained network: 256 × 256 pixels. However, the images of lung radiographs (both from portable and fixed devices) are of larger size and proportions. This implies that, during the resizing and compression of the information, much of it is being deformed and/or lost.

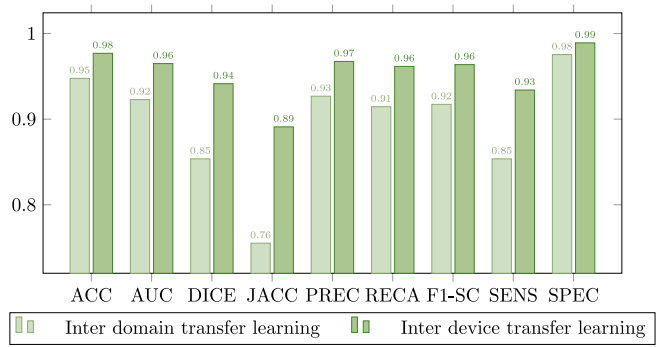

**Fig. 11.** Comparison between the model trained with common chest X-ray radiographs and the model adapted to portable devices for images from patients diagnosed with lung afflictions similar to COVID-19 (such as pneumonia) but not COVID-19 related.

**Table 7**
Mean time and standard deviation in milliseconds for each knowledge transfer stage.

| Transfer stage | Mean | | Standard deviation | |
|---|---|---|---|---|
| | MRI to General | General to Portable | MRI to General | General to Portable |
| Training | 35,413.26 | 1,857.17 | 42.92 | 13.17 |
| Validation | 3,910.10 | 353.18 | 17.17 | 11.74 |

Finally, in the Subsection 4.3 we see how the real load of the knowledge transfer falls on the first stage. As previously commented, this is the stage where we are really transforming the filters learned by the pretrained network in the MRI domain to the chest radiographs one. The second stage helps to teach the network how to support the possible adverse scenarios that the methodology may encounter with the image modality coming from portable devices. These shorter times in the second stage, at first, could be attributed exclusively to a lower number of images per epoch; but if we look at the Figs. 7 and 8 we see that the number of epochs needed at most for the second stage of transfer (remember that our methodology is based on an early stopping when no improvement is achieved for a given number of epochs) is also significantly lower. This indicates that not only this amount of time is the result of a lower dataset, but also that the system converges earlier than in the first stage.

## 6. Conclusions

In this work, we have proposed a system with the purpose of segmenting lung regions in thoracic radiographs, especially aimed at those obtained by portable X-ray devices in adverse conditions and with a limited supply of images. These devices, which represent an alternative to fixed devices to reduce the risk of cross-contamination in the diagnosis of diseases such as COVID-19, are critical for medical emergency situations, so methodologies aimed to help in the diagnostic process that are functional with these images are critical. To solve the problem of poor image quality due to the capture conditions and devices themselves, we propose a fully automatic methodology based on two stages of transfer learning. A first stage based on knowledge transfer from a domain similar to radiographs and trained with a large number of images (ensuring its robustness) to common chest radiographs obtained from different public sources and a second stage, in which knowledge is refined to adapt it to specific radiographs of a dataset obtained in adverse conditions in the clinical practice during the pandemic.

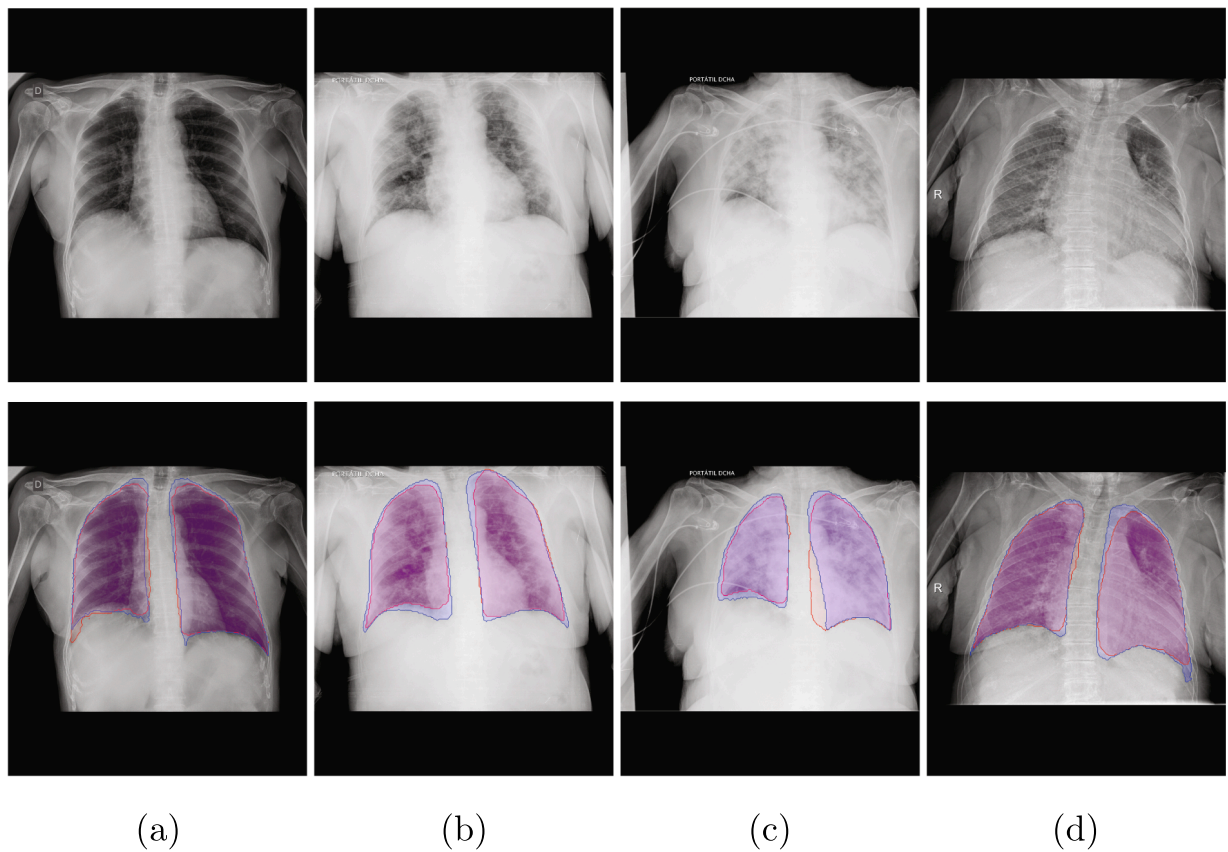

**Fig. 12.** Sample images of the network output from the inter domain transfer learning stage (red) and inter device learning stage (blue). (a): Normal, (b): COVID-19, (c) & (d): Non-covid lung pathologies. First row: original image, Second row: comparison between models.

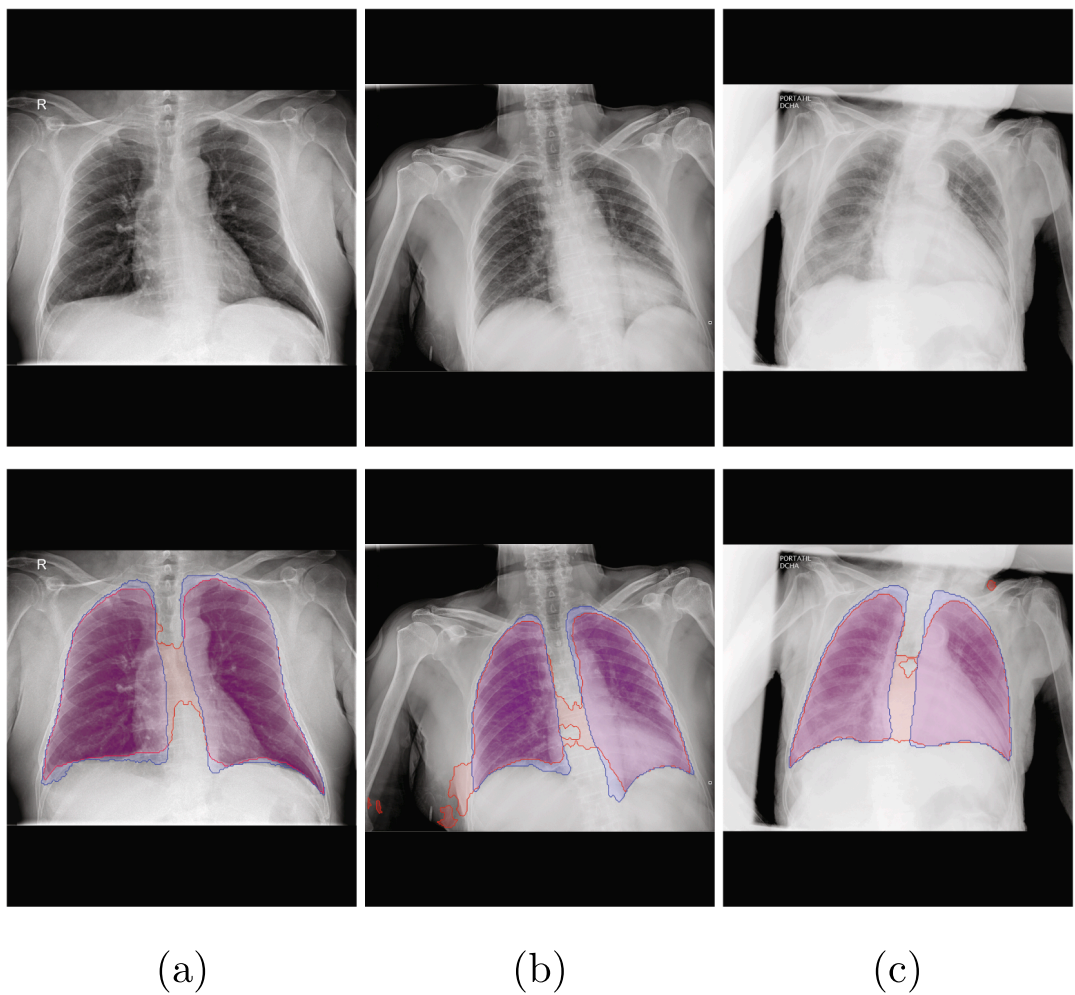

**Fig. 13.** Sample images with unwanted artifacts and formations from the inter domain transfer learning stage (red) and that same output from the inter device type learning stage (blue). (a): Normal, (b): COVID-19, (c): Non-covid lung pathologies. First row: original image, Second row: comparison between models.

As we have shown in the metrics of the results and in the discussion, while the first stage of transfer learning allows the system to acquire the knowledge bases of the domain to generate an initial segmentation, the second stage of knowledge transfer to the particular domain manages to refine satisfactorily the obtained segmentations even with a limited set of samples. This second stage of transfer learning allows not only to better estimate the pulmonary region, but also to eliminate various artifacts resulting from the lower sample quality present in the images from portable devices.

Thus, as a final result of this work, we have successfully obtained a fully automatic methodology based on deep methodologies, using a limited number of images from portable devices and capable of working with these images in a robust and consistent way, regardless of the image quality and capture conditions.

As future work, it would be interesting to study mechanisms to adapt the network input resolution so that it could support variable input sizes (in addition to study the performance difference between both proposals) to solve the border degradation in the segmentation product of the rescaling of the images. Another aspect that would be desirable to improve is the network that was used as a basis for knowledge transfer. This network is receiving as input an image of resolution $256 \times 256 \times 3$. However, the pulmonary radiography images we use only have useful information in grayscale (ergo the information is replicated along the three input channels). It would be interesting to explore other works as foundation that, like our images, employ a network with single-channel input to make the training more efficient and possibly improve its generalization capacity (by reducing the overall size of the network).

Another known problem of transfer learning with the technique we use is the abrupt changes of the gradient during the training that can cause the degradation of features already learnt by the network during the pretraining. An alternative technique for knowledge transfer is the addition of new layers at the end of a pre-trained network and freezing the weights of the original layers. By doing so, the network would be extended with a sort of "domain translator module". Thus, the feature extraction mechanism of the original network would be kept static (its weights would not be altered during training) and, consequently, the features learned during the basic training would be fully preserved.

On the other hand, given the positive results obtained in the application of this methodology, we see that, in fact, the features present in MRI image of cerebral glioma are reusable in the field of lung region segmentation in portable chest X-rays. Another interesting future work would consist in the so-called "deep feature analysis", which would allow to study the common features learned by the network in both domains and thus help to better understand and improve the present and future clinical diagnostic support systems. Additionally, given that all the images analyzed in the portable dataset come almost no patients with implants or foreign objects that could leave artifacts in the chest radiographs, it would be interesting to study the impact of these devices on the capabilities of the system to correctly infer the lung region, as well as (connecting to the previous topic) the effect on these artifacts on the features the networks deems relevant to detect them.

**Funding**

This research was funded by Instituto de Salud Carlos III, Government of Spain, DTS18/00136 research project; Ministerio de Ciencia e Innovación y Universidades, Government of Spain, RTI2018-095894-B-I00 research project, Ayudas para la formación de profesorado universitario (FPU), grant Ref. FPU18/02271; Ministerio de Ciencia e Innovación, Government of Spain through the research project with reference PID2019-108435RB-I00; Consellería de Cultura, Educación e Universidade, Xunta de Galicia, Grupos de Referencia Competitiva, grant Ref. ED431C 2020/24; Axencia Galega de Innovación (GAIN), Xunta de Galicia, grant Ref. IN845D 2020/38; CITIC, as Research Center accredited by Galician University System, is funded by "Consellería de Cultura, Educación e Universidade from Xunta de Galicia", supported in an 80% through ERDF Funds, ERDF Operational Programme Galicia 2014-2020, and the remaining 20% by "Secretaría Xeral de Universidades" (Grant ED431G 2019/01).

**CRediT authorship contribution statement**

**Plácido L. Vidal:** Conceptualization, Methodology, Software, Formal analysis, Investigation, Data curation, Writing - original draft, Writing - review & editing, Visualization. **Joaquim de Moura:** Conceptualization, Validation, Investigation, Data curation, Writing - review & editing, Supervision, Project administration. **Jorge Novo:** Validation, Investigation, Data curation, Writing - review & editing, Supervision, Project administration, Funding acquisition. **Marcos Ortega:** Validation, Investigation, Data curation, Writing - review & editing, Supervision, Project administration, Funding acquisition.

**Declaration of Competing Interest**

The authors declare that they have no known competing financial interests or personal relationships that could have appeared to influence the work reported in this paper.